\documentclass{svproc}
\usepackage{url}
\usepackage{epsfig}

\begin{document}
\mainmatter              
\title{Magnetic resonance imaging assessment of the suitability and consistency of radiotherapy treatment positioning achieved using intra-oral stents}
\titlerunning{MRI imaging of intra-oral stents}  

\author{Tanya Kairn\inst{1,2,3} \and Philip Chan\inst{1,2} \and Benjamin Chua\inst{1,2} \and Susannah Cleland\inst{1,4}\footnote{Susannah Cleland, present address: Radiation Oncology Princess Alexandra Hospital Raymond Terrace, South Brisbane, QLD, 4101, Australia} \and Jodi Dawes\inst{1} \and Lizbeth Kenny\inst{1,2} \and Charles Y. Lin\inst{1,2} \and William R. McDowall\inst{1} \and Tania Poroa\inst{1,4} \and Scott B. Crowe\inst{1,2,3,4}}

\authorrunning{Tanya Kairn et al.}
\tocauthor{Tanya Kairn, Philip Chan, Benjamin Chua, Susannah Cleland, Jodi Dawes, Lizbeth Kenny, Charles Y. Lin, William R. McDowall, Tania Poroa, and Scott B. Crowe}

\institute{Royal Brisbane and Women's Hospital, Brisbane, Qld, Australia\\
\and
University of Queensland, Brisbane, Qld, Australia
\and
Queensland University of Technology, Brisbane, Qld, Australia
\and
Herston Biofabrication Institute, Metro North Hospital and Health Service, Brisbane, Qld, Australia\\
\email{t.kairn@gmail.com}}

\maketitle              

\begin{abstract}
As head-and-neck radiotherapy treatments grow more complex and precise, it becomes increasingly important to assess the anatomical separations that can be achieved using intra-oral stents. A series of twenty T2-weighted turbo spin echo magnetic resonance images (MRI) were acquired of one healthy participant, with a range of different wax and 3D printed intra-oral stents in situ. The resulting measurements showed that a 3D printed modular stent containing hard polylactic acid (PLA) and flexible thermoplastic polyurethane (TPU) components made the largest and most reproducible separation between the cheeks (70.8 +/- 0.3 mm), two hard PLA stents designed to exactly fit the participant’s teeth produced the poorest positioning reproducibility (standard deviations of up to 3 mm between a range of landmarks measured in repeated images). Most stents were described as “comfortable” although the wax stents left small pieces of wax attached to the teeth after use. This MRI based comparison demonstrated that the materials and designs used for intra-oral stents can have substantial effects on the level of anatomical separation and positioning reproducibility that they produce. 
\end{abstract}

\section{Introduction}

The lips, tongue, cheeks (buccal mucosa) and palate are all comparatively radiation sensitive structures that can affect quality of life if damaged during radiotherapy treatments of nearby cancers \cite{parliament2004,navran2019}. Given the proximity and flexibility of many of these structures, it is often desirable and comparatively straightforward to achieve a physical separation though the use of intra-oral stents \cite{appendino2019,alves2021}. As head-and-neck radiotherapy treatments grow more complex and precise \cite{parliament2004,navran2019,stieler2011}, it becomes increasingly important to assess the positioning that can be achieved using intra-oral stents.

Previous studies have investigated stents made by dentists using acrylic resins \cite{appendino2019,verrone2012} or made in-house in radiotherapy departments using wax \cite{norfadilah2017,cleland2021} or fused deposition modelling (FDM) 3D printing techniques \cite{cleland2021,hong2019}. Due to the undesirability of repeatedly imaging patients using ionising radiation, such as via repetition of planning computed tomography (CT) acquisitions with different stents, stent evaluations have generally used CTs of different patients with specific stents \cite{appendino2019,verrone2012,hong2019} with comparison studies using phantoms rather than human subjects \cite{cleland2021}.

Magnetic resonance imaging (MRI) is a comparatively safe, non-ionising modality that produces images with improved soft-tissue contrast compared to CT \cite{schmidt2015}. This study therefore used MRI to evaluate the anatomical displacements achieved by six different intra-oral stents when used by one healthy volunteer, to evaluate the comparative suitability of these stents for use in radiotherapy treatments for head-and-neck cancer.

\section{Method}

This study evaluated three pairs of oral stents that were constructed via three different methods. One pair of stents was made from wax, one replica pair was created by 3D printing geometric copies of the wax stents, and the third pair was constructed using pre-prepared pieces from an in-house system of modular 3D printed components. All stents were designed to maintain a mandible-maxilla separation of approximately 20 mm, while also depressing the tongue away from the palate and displacing one or both cheeks away from the tongue and teeth. In each pair, a ``one-sided'' stent was designed to displace one cheek and a ``two-sided'' stent was designed to displace both cheeks.

The first pair of stents was constructed using the established method \cite{cleland2021,lee2019} of forming wax around a hollow plastic tube (breathing aperture) and then molding the wax to fit the subject's teeth. This pair of stents was CT scanned using a Siemens Somatom Confidence scanner (Siemens AG, Erlangen, Germany), using high-resolution scanning parameters described previously \cite{kairn2021}. The resulting images were converted into stereolithography files via threshold segmentation and into gcode 3D printing instructions respectively using 3DSlicer \cite{fedorov2012} and Cura (Ultimaker BV, Geldermalsen, Netherlands) software, so that 3D printed replicas of the two stents could be created using an Ender 5 3D printer (Creality 3D, Shenzhen, China). The 3D printed replicas were printed from a polylactic acid (PLA) filament using a gyroid infill pattern and a 10\% in-fill density to achieve mechanically robust construction at a sufficiently low density to avoid producing perturbations or dose build-up if used in a radiotherapy beam. 

Our in-house modular intra-oral stent system was used to construct the third pair of stents. This system allows intra-oral stents to be created using a small library of pre-prepared pieces. Each stent used in this work consisted of a central main piece 3D printed from PLA at 30\% in-fill  0.33$\pm$0.05 g/cm$^3$), to achieve the required mouth-opening and tongue depressing effects, plus one or two cheek-displacement pieces 3D printed from flexible thermoplastic polyurethane (TPU) with different sections printed at 12 to 30\% in-fill (0.0010$\pm$0.0008 to 0.27$\pm$0.04 g/cm$^3$), to achieve cushioned and comfortable cheek displacement with sturdy connections to each main piece.

The anatomical effects of the three pairs of oral stents were evaluated via a series of twenty T2-weighted turbo spin echo MRI scans of one healthy participant, with and without the various stents in situ. Imaging with each stent was repeated two to three times, with a delay of at least 60 minutes between repeat scans of each stent. Imaging was performed using a Siemens Prisma 3T MRI (Siemens AG, Erlangen, Germany). Acquisition of each MRI image took approximately four minutes. 

Acquisition of this MRI series was approved by the Royal Brisbane and Women's Hospital Human Research Ethics Committee (RBWH HREC, EC00172) and this study was completed in accordance with the ethical standards of the  RBWH HREC and the 1964 Helsinki declaration and its later amendments.

The resulting images were evaluated using measuring tools available in the MIM Maestro software package (MIM Software Inc, Cleveland, USA). Sagittal views were used to measure the length of tongue depressed (the flattened region of the upper surface of the tongue) the jaw-opening distance (distance between visible soft tissue on the mandible and maxilla sides of the mouth) and the maximum distance from tongue to palate, in all MRI images with oral stents. Coronal views were used to measure the maximum cheek-to-cheek distance (the maximum distance between the left and right inner cheek walls), in all MRI images.

\section{Results}

\begin{figure}[hbtp]
\vspace{-0.5cm}
\begin{center}
\includegraphics[width=\linewidth]{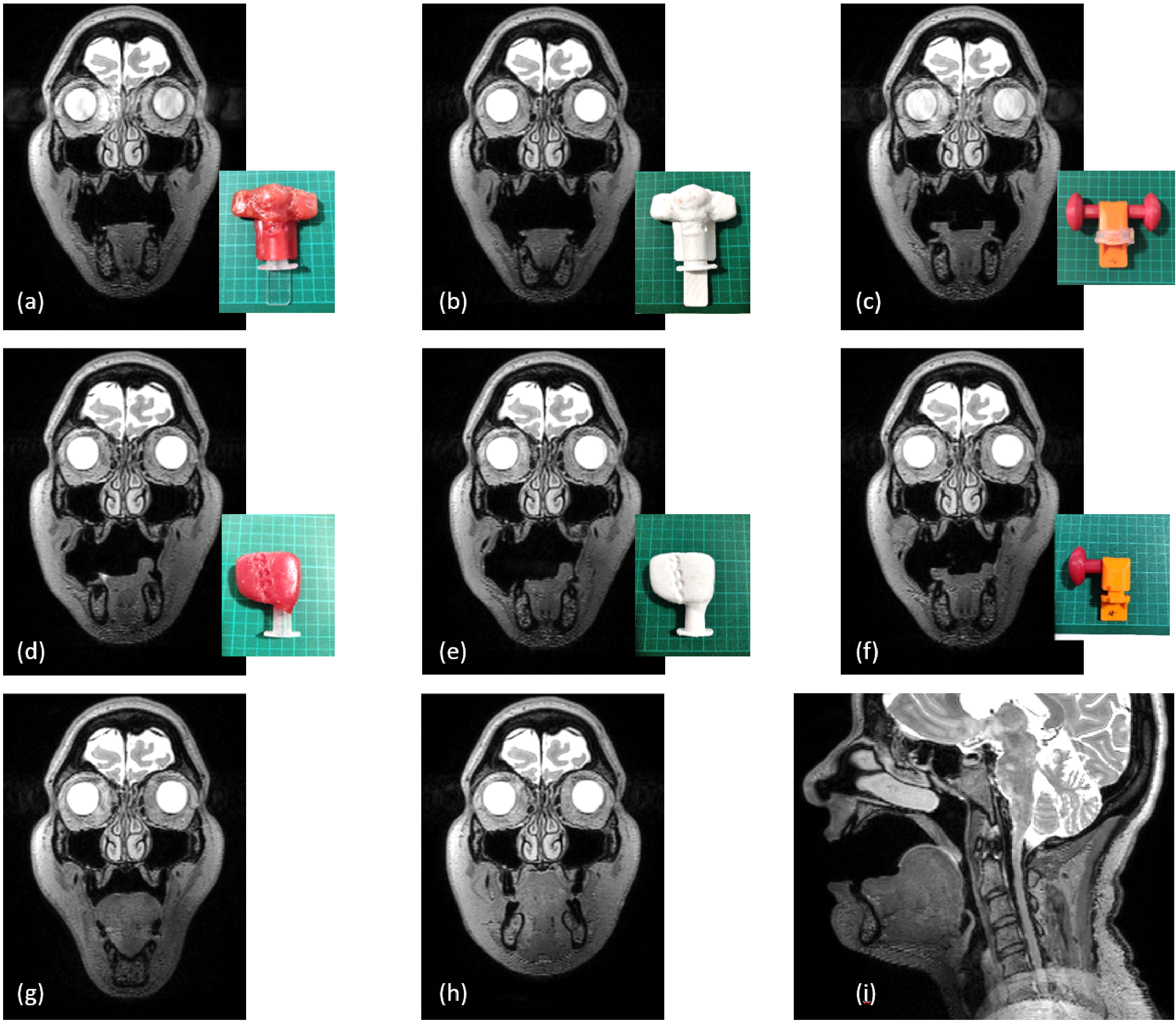}
\caption{(a)-(h) coronal and (i) sagittal slices through MRI images of participant with and without intra-oral stents in situ: (a) two-sided wax stent, (b) two sided 3D printed replica of wax stent, (c) two sided modular stent, (d) one-sided wax stent, (e) one sided 3D printed replica of wax stent, (f) one sided modular stent, (g) no stent with mouth open, (h) no stent with mouth closed, (i) two-sided modular stent. Insets show stents used to acquire the corresponding coronal views.}
\label{fig:1}
\end{center}
\vspace{-0.5cm}
\end{figure}

Figures \ref{fig:1}(a)-(f) show example coronal slices from the MRI images of the participant with intra-oral stents in situ, with photographs of the corresponding stents included as insets. Figures \ref{fig:1}(g) and (h) show example coronal images of the participant without any intra-oral stent, with their mouth respectively opened and closed. Figure \ref{fig:1}(i) shows an example sagittal image indicating the mouth-opening and tongue-depressing effect of one of the modular stents.

None of the materials in the intra-oral stents (wax, PLA and TPU) produced a resonance signal, so all stents appeared black in figures \ref{fig:1}(a)-(f). However, the soft-tissue contrast seen in the images allowed straightforward measurement of a range of anatomical landmarks, clearly indicating the separations between different tissues that were achieved by each stent. These measurement results are summarised in table \ref{tab:1}.

The measurements listed in table \ref{tab:1} indicate that the 3D printed modular stent containing hard (PLA) and flexible (TPU) components achieved the largest and most reproducible separation between the cheeks (78.0 $\pm$ 0.3 mm). By contrast, the hard PLA stents that were designed to replicate the wax stents and thereby exactly fit the participant’s teeth produced the poorest positioning reproducibility, with standard deviations of up to 0.3 cm between a range of landmarks measured in repeated images). 

The participant remarked that the wax stents felt slightly malleable and ``clingy'', leading to a feeling of confidence that the stents were securely positioned. However, reproducible positioning was difficult to achieve due to the ease of simply biting new tooth positions into the wax. The participant also found that small pieces of wax remained adhered to the teeth every time a wax stent was removed, indicating degradation of the wax stents by repeated use.

The 3D printed replicas of the wax stents were described by the participant as the least comfortable, because their smooth and hard surfaces felt slippery and unstable in the mouth, leading to a feeling that the participant needed to grip the stents with their jaw to achieve stable positioning.

The 3D printed modular stent was described as the most comfortable, despite not being fitted to the participant's oral anatomy. The simple groove on the top of the stent allowed reproducible positioning of the incisors and the cheek-displacement pieces fitted comfortably over the premolars, keeping the stent fixed in place without effort.

While all stents had breathing holes that the participant found ``usable'', the participant reported feeling comfortable enough with all intra-oral stents to allow relaxed nasal breathing for most of the duration of the imaging session.

\begin{table}[htbp]

\setlength{\tabcolsep}{5pt}
\caption{Measured distances (mm) between oral landmarks observed in neighbouring slices of repeated MRI images with and without the various types of intra-oral stent.}
\vspace{-0.3cm}
\label{tab:1}
\begin{center}
\begin{tabular}{ccccc}
\hline
Stent type      &Length of tongue&Jaw opening   &Max distance   &Max distance\\
                &depressed      &distance       &tongue to palate&cheek to cheek\\
\hline\rule{0pt}{12pt}
Wax 2-sided     &26$\pm$1    &26.4$\pm$0.8  &33.4$\pm$0.4  &75.6$\pm$0.5\\
Replica 2-sided &21$\pm$3    &30$\pm$2      &35.7$\pm$0.7  &76.4$\pm$0.8\\
Modular 2-sided &22$\pm$1    &27.2$\pm$0.7  &37.4$\pm$0.3  &78.0$\pm$0.3\\
Wax 1-sided     &23.2$\pm$0.5&19$\pm$2      &33.7$\pm$0.8  &68$\pm$1\\
Replica 1-sided &25$\pm$2    &22.8$\pm$0.5  &36.5$\pm$0.3  &66.7$\pm$0.5\\
Modular 1-sided &20$\pm$1    &27.3$\pm$0.5  &35.7$\pm$0.7  &65.9$\pm$0.7\\
Open mouth      &N/A         &33$\pm$3      &33$\pm$3      &53$\pm$1\\
Closed mouth    &N/A         &N/A           &N/A           &53$\pm$3\\[2pt]
\hline
\end{tabular}
\end{center}
\vspace{-1cm}
\end{table}

\section{Discussion}

Results summarised in figure \ref{fig:1} and table \ref{tab:1} indicate that all six stents were able to achieve mouth-opening, tongue-depressing and cheek-displacing effects. The participant described most of the stents as ``comfortable'' although the wax stents were degraded by repeated use and the 3D printed replicas of the wax stent were described as the least comfortable. Difficulty maintaining the 3D printed replicas in a consistent position also led to variable measurement results, as indicated by standard deviations in table \ref{tab:1}.

The comfort, stability and reproducibility of results from the 3D printed modular stents suggested that these pre-prepared customisable stents may produce optimal results for head-and-neck radiotherapy patients despite not being molded to fit the patients' specific oral anatomy. The positive results achieved with these modular stents, which incorporated slightly flexible components printed from TPU using an increased in-fill density, also indicate a potential method to reduce the feelings of slipperiness and instability observed with the 3D printed replicas of wax stents, by similarly printing with TPU at an increased in-fill density.

\section{Conclusion}

Results of this study showed how MRI imaging can allow useful measurements and comparisons of oral tissue separations to be made, so that the suitability and consistency of different stent designs for use in radiotherapy treatments for head-and-neck cancer can be assessed. Evidently, the materials and designs used for intra-oral stents can have substantial effects on the level of anatomical separation and positioning reproducibility that they produce. These results suggest that when 3D printing intra-oral stents, the careful selection of materials, densities and design features has the potential to produce intra-oral stents that achieve improved comfort and stability, even when not specifically shaped to fit the particular patient's oral anatomy. 

\section*{Acknowledgements}
Contributions to this work from Susannah Cleland, Jenna Luscombe, Tania Poroa and Scott B. Crowe were supported by a Metro North Hospital and Health Service funded Herston Biofabrication Institute Programme Grant. MRI image acquisition was made possible by a grant from the Herston Imaging Research Facility – Project Support Scheme 2020.

\end{document}